\documentclass[12pt]{article}
\usepackage{putex}
\usepackage{graphicx}

\def\arctanh{\mop{arctanh}}

\begin{document}

\preprint{PUPT-2340}

\institution{PU}{Joseph Henry Laboratories, Princeton University, Princeton, NJ 08544, USA}

\title{Symmetry constraints on generalizations of Bjorken flow}

\authors{Steven S. Gubser\footnote{e-mail: {\tt ssgubser@Princeton.EDU}}}

\abstract{I explain a generalization of Bjorken flow where the medium has finite transverse size and expands both radially and along the beam axis.  If one assumes that the equations of viscous hydrodynamics can be used, with $p=\epsilon/3$ and zero bulk viscosity, then the flow I describe can be developed into an exact solution of the relativistic Navier-Stokes equations.  The local four-velocity in the flow is entirely determined by the assumption of symmetry under a subgroup of the conformal group.}

\date{June 2010}

\maketitle

\tableofcontents

\section{Introduction}

Bjorken flow \cite{Bjorken:1982qr} is an attempt to describe the average motion of partons after a collision of heavy ions.  Three assumptions that go into the treatment of \cite{Bjorken:1982qr} are approximate boost-invariance along the beamline near mid-rapidity, translation invariance in the transverse plane, and rotation invariance in that plane.  If Minkowski space is parametrized\footnote{More properly, the coordinates \eno{BjMink} cover a wedge of ${\bf R}^{3,1}$ which is the causal future of the collision plane.} as
 \eqn{BjMink}{
  ds^2 = -d\tau^2 + \tau^2 d\eta^2 + dx_\perp^2 + x_\perp^2 d\phi^2 \,,
 }
then translation and rotation invariance in the transverse plane imply that nothing can depend on $x_\perp$ or $\phi$, while boost-invariance (if exact) implies that nothing depends on $\eta$.  Together with invariance under reflections $\eta \to -\eta$, these symmetries imply that the local four-velocity vector is $u^\mu = (1,0,0,0)$ in the $(\tau,\eta,x_\perp,\phi)$ coordinate system.  One does not need any information about the equation of state to determine the four-velocity profile.  Symmetry considerations alone fix it.  The medium need not even be equilibrated for $u^\mu = (1,0,0,0)$ to be a well-motivated choice for the average four-velocity in the mid-rapidity region.  If the medium is equilibrated and has equation of state $p = \epsilon/3$, then the energy density in the local rest frame scales as $1/\tau^{4/3}$.

Translation invariance in the transverse plane is, of course, highly unrealistic since the nucleus is only about $13\,{\rm fm}$ across.  Treatments based on the Bjorken picture often assume that the medium has no average local velocity in the $x_\perp$ direction (radial flow) until after it thermalizes.  This is probably wrong, even for perfectly central collisions, and one might worry that it significantly distorts the subsequent hydrodynamical flows on which much of heavy-ion phenomenology depends.  Indications that the absence of initial radial flow could be problematic can be found, for example, in \cite{Kolb:2002ve,Heinz:2004qz}; see also \cite{Lisa:2008gf,Broniowski:2008ee,Pratt:2008qv}.  The question naturally arises, can we estimate in some way deviations from the Bjorken picture, based on the finite size of the colliding nuclei, which lead to non-zero $u^{x_\perp}$?  The aim of this paper is to present a generalization of Bjorken flow which does exactly that.

The approach I will follow is based mostly on symmetry considerations, and it requires that the initial state and the dynamics of the collision perfectly respect relativistic conformal invariance.  Also, I am limited to perfectly central collisions.  Obviously, these assumptions are by no means exactly satisfied in real collisions.  However, Quantum Chromodynamics (QCD) processes well above the confinement scale are close to conformally invariant because the coupling runs only logarithmically with scale.  So conformally invariant dynamics, especially in early stages of the collisions, is an interesting start point.  The key input which allows me to extract a velocity profile solely from symmetry considerations is the assumption that translations in the transverse plane can be replaced by certain special conformal transformations which, in combination with rotations around the beamline, form an $SO(3)$ subgroup of the full conformal group.  The significance of this $SO(3)$ subgroup was previously noted in calculations \cite{Gubser:2008pc} based on the gauge-string duality \cite{Maldacena:1997re,Gubser:1998bc,Witten:1998qj}.  Indeed, ideas in \cite{Gubser:2008pc,Gubser:2009sx} (see also \cite{Kovchegov:2007pq,Grumiller:2008va}) strongly underlie the proposal in this paper.  But I do not depend on any of the dynamical information that the gauge-string duality provides.  This is in contrast to \cite{Janik:2005zt}, where a gauge-string dual description of Bjorken flow was found by constructing an approximate solution to Einstein's equations in $AdS_5$.  The main calculations in this paper rely only on symmetries and hydrodynamics. 

The organization of the rest of the paper is as follows.  In section~\ref{SUBGROUP} I will explain the aspects of conformal symmetry that I am going to use.  In section~\ref{VELOCITY} I use conformal symmetry to pick out a special four-velocity profile.  In section~\ref{INVISCID} I will use conformal symmetry again to help find an energy density which, along with the special four-velocity profile, solves the equations of inviscid, conformally invariant relativistic hydrodynamics.  In section~\ref{VISCOUS} I extend the result to the case of non-zero shear viscosity.  The exact solution of the Navier-Stokes equations that I find generalizes the result $\epsilon \propto 1/\tau^{4/3}$ for Bjorken flow.  Unless the viscosity is identically zero, there is a pathology in the solution at very early times and/or very large radius.  This pathology can be understood as the breakdown of a hydrodynamical description.  In section~\ref{EARLY} I characterize the colliding shock waves which respect the same conformal symmetries as the special four-velocity profile after the collision.  In section~\ref{NUMBERS} I use results on total multiplicity to estimate numerical parameters in the hydrodynamical flows found in previous subsections.  In section~\ref{COVERING} I explain how some of the features of the flow are more transparent when one maps it to $S^3 \times {\bf R}$, which is the covering space of Minkowski space, ${\bf R}^{3,1}$.  Readers wishing to skip the motivations and technical detail can find a brief summary of the main results in section~\ref{SUMMARY}.

\section{An $SO(3)$ subgroup of the conformal group}
\label{SUBGROUP}

The conformal group in four dimensions is an extension of the Poincar\'e group $ISO(3,1)$ to $SO(4,2)$.  The generators of the Lie algebra underlying $SO(4,2)$ are (essentially in the notation of \cite{Ginsparg:1988ui}):
 \begin{itemize}
  \item Translations: $\xi^\mu = a^\mu$ where $a^\mu$ is constant.
  \item Rotations and boosts: $\xi^\mu = \omega^\mu{}_\nu x^\nu$ where $\omega_{\mu\nu}$ is constant and anti-symmetric.
  \item Scale transformations: $\xi^\mu = x^\mu$.
  \item Special conformal transformations: $\xi^\mu = x^\nu x_\nu b^\mu - 2 b^\nu x_\nu x^\mu$ where $b^\mu$ is constant.
 \end{itemize}
Obviously, the first two classes of symmetries belong to the Poincar\'e group, $ISO(3,1)$.  Geometrically, they are Killing vectors, satisfying ${\cal L}_\xi g_{\mu\nu} = 0$ where ${\cal L}_\xi$ denotes the Lie derivative and $g_{\mu\nu}$ is the standard metric of ${\bf R}^{3,1}$ with mostly plus signature.\footnote{Readers unfamiliar with Lie derivatives can understand the main points by interpreting ${\cal L}_\xi$ as an operator which measures how a tensor changes due to a coordinate transformation $x^\mu \to x^\mu + \xi^\mu$, treated to linear order in $\xi^\mu$.  The main properties of ${\cal L}_\xi$ are linearity, the Leibniz rule for products, and the following definitions:
 \eqn{LieDefs}{\seqalign{\span\TL & \span\TR &\qquad\qquad\span\TT}{
  {\cal L}_\xi \phi &\equiv \xi^\mu {\partial\phi \over \partial x^\mu} & 
    for scalars $\phi$  \cr
  {\cal L}_\xi u^\mu &\equiv \xi^\lambda {\partial u^\mu \over \partial x^\lambda} - 
    u^\lambda {\partial\xi^\mu \over \partial x^\lambda} & for vectors $u^\mu$  \cr
  {\cal L}_\xi \zeta_\mu &\equiv \xi^\lambda {\partial\zeta_\mu \over \partial x^\lambda} + 
    \zeta_\lambda {\partial\xi^\lambda \over \partial x^\mu} & for $1$-forms $\zeta_\mu$.
 }}
While ${\cal L}_\xi$ is usually defined in terms of partial derivatives $\partial/\partial x^\mu$, one may instead use the covariant derivative $\nabla_\mu$, defined in terms of the Christoffel connection, without changing the results.  A particularly useful result is ${\cal L}_\xi g_{\mu\nu} = \nabla_\mu \xi_\nu + \nabla_\nu \xi_\mu$ for the metric tensor $g_{\mu\nu}$.

It is also good to note that ${\cal L}_\xi u^\mu = -{\cal L}_u \xi^\mu$.  One often denotes ${\cal L}_\xi u$ by $[\xi,u]$, also known as the Lie bracket, or simply as the commutator of $\xi$ and $u$.}  The vector fields associated with scale transformations and special conformal transformations are not Killing vectors, but rather {\it conformal} Killing vectors, satisfying
 \eqn{CKV}{
  {\cal L}_\xi g_{\mu\nu} = {1 \over 2} (\nabla_\lambda \xi^\lambda) g_{\mu\nu} \,.
 }
Often, I will follow conventions of differential geometry in specifying a vector in terms of a combination of derivatives: for example, $\xi = a^\mu {\partial \over \partial x^\mu}$ corresponds to $\xi^\mu = a^\mu$.

As reviewed in the introduction, the symmetries of Bjorken flow are:
 \begin{enumerate}
  \item Rotation invariance around the beamline, $\xi = x^1 {\partial \over \partial x^2} - x^2 {\partial \over \partial x^1}$.  Of course, this is an exact symmetry only in the vanishingly rare case of a perfectly head-on collision. 
  \item Translation invariance in the transverse plane, $\xi = {\partial \over \partial x^1}$ and $\xi = {\partial \over \partial x^2}$.  This is the symmetry I am most interested in relaxing, since it forbids radial flow in the transverse plane.
  \item Boost invariance along the beamline, $\xi = x^3 {\partial \over \partial t} + t {\partial \over \partial x^3}$.  This symmetry is the key feature of Bjorken's treatment.  It is based on a model where the Lorentz-flattened nuclei largely pass through one another and leave behind a medium in the wedge $|x^3| < t$.  The boost invariance is supposed to hold only in some neighborhood of mid-rapidity---which is to say, for $|x^3/t|$ not too close to $1$. 
 \end{enumerate}
In order to show that these three symmetries, together with symmetry under the ${\bf Z}_2$ group reflecting $x^3 \to -x^3$, completely fix the local four-velocity $u^\mu$, the equations we solve are $[\xi,u] = 0$ where $\xi$ is any of the symmetries in the numbered list above.  The only solution consistent with the ${\bf Z}_2$ symmetry is $u = {\partial \over \partial\tau}$, where we pass to coordinates $(\tau,\eta,x_\perp,\phi)$ defined as
 \eqn[c]{SemiPolar}{
  \tau = \sqrt{t^2 - (x^3)^2} \qquad \eta = \arctanh {x^3 \over t}  \cr
  x_\perp = \sqrt{(x^1)^2 + (x^2)^2} \qquad \phi = \arctan {x^2 \over x^1} \,.
 }
A formal phrasing of what makes $u = {\partial \over \partial\tau}$ special is that it is the only timelike unit vector which is invariant under the $ISO(2)$ symmetry of the transverse $x^1$-$x^2$ plane, the $SO(1,1)$ boost symmetry in the $x^3$ direction, and the ${\bf Z}_2$ symmetry that reflects through the collision plane.\footnote{Relaxing our insistence on the ${\bf Z}_2$ symmetry while preserving the continuous symmetries would allow non-zero $u^\eta$.  This is a mild generalization because one could perform an overall boost to restore the ${\bf Z}_2$ symmetry.}

The $ISO(2) \times SO(1,1)$ symmetry also implies that $\epsilon$, the energy density in the local rest frame of the medium, can depend only on $\tau$.  But in order to actually determine the functional dependence $\epsilon(\tau)$, one needs much more information.  For example, if hydrodynamics is assumed to be valid, then one needs conservation of the stress-energy tensor, constitutive relations, the equation of state, and a choice of shear and bulk viscosity.  

I am interested in modifying the symmetry constraints in order to accommodate finite transverse size and radial flow.  After some thought starting from the treatment \cite{Gubser:2008pc,Gubser:2009sx} of colliding shocks in $AdS_5$,\footnote{Already in \cite{Gubser:2008pc,Gubser:2009sx} the hope was expressed that heavy-ion collisions might be approximately invariant under an $O(3)$ symmetry (or an $O(2)$ subgroup for off-center collisions) generated as explained here.  Failure of motivation and insight, in some combination, prevented me from actually writing down the key equation \eno{uSatisfies} for almost two years.} I concluded that the way to proceed is to replace translation invariance by invariance under
 \eqn{xiDefs}{
  \xi_{i4} = {\partial \over \partial x^i} + q^2 
    \left[ 2 x^i x^\mu {\partial \over \partial x^\mu} - 
       x^\mu x_\mu {\partial \over \partial x^i} \right] \,,
 }
where $i=1$ or $2$ and $q$ is a quantity with dimensions of inverse length.  The vector $\xi_{14}$ (and analogously $\xi_{24}$) is so named because it is inherited from a symmetry of $AdS_5$, embedded in ${\bf R}^{4,2}$ by the equation\footnote{More properly, $AdS_5$ is the universal cover of the solution space to \eno{AdSEqn}.}
 \eqn{AdSEqn}{
  -(X^{-1})^2 - (X^0)^2 + (X^1)^2 + (X^2)^2 + (X^3)^2 + (X^4)^2 = -1 \,,
 }
and this symmetry acts on $X^1$ and $X^4$ as an ordinary rotation.  Even without reference to $AdS_5$, \eno{xiDefs} is a more obvious thing to try than any other conformal transformation outside of $ISO(3,1)$, because it is a combination, for $i=1$, of translation by $a^\mu = \delta^\mu_1$ with a special conformal transformation with $b^\mu \propto \delta^\mu_1$.  It is easily checked that the commutator of $\xi_{14}$ and $\xi_{24}$ is proportional to ${\partial \over \partial\phi}$, which is rotation around the beam axis.  Indeed, $\xi_{14}$, $\xi_{24}$, and ${\partial \over \partial\phi}$ generate an $SO(3)$ subgroup of $SO(4,2)$ which commutes with the $SO(1,1)$ subgroup generated by $\xi = x^3 {\partial \over \partial t} + t {\partial \over \partial x^3}$.  The particular choice of the $SO(3)$ is controlled by the parameter $q$, and as $q \to 0$, the $SO(3)$ degenerates to the $ISO(2)$ symmetry of Bjorken flow.  Let's call the group generated by $\xi_{i4}$ and $\xi_\phi$ $SO(3)_q$ to remind ourselves that it depends on $q$ in an interesting way.

In order to consider $SO(3)_q$ symmetry further, it will be convenient to recast one of the conformal Killing vectors \eno{xiDefs} in terms of the coordinates \eno{SemiPolar}:
 \eqn{zetaDef}{
  \zeta \equiv \xi_{14} = 2 q^2 \tau x_\perp \cos\phi {\partial \over \partial\tau} + 
    (1 + q^2 \tau^2 + q^2 x_\perp^2) \cos\phi {\partial \over \partial x_\perp} - 
    {1 + q^2 \tau^2 - q^2 x_\perp^2 \over x_\perp} \sin\phi {\partial \over \partial\phi}
      \,.
 }

\section{A special four-velocity profile}
\label{VELOCITY}

A definite problem is now almost formulated: I want to find a four-velocity profile in ${\bf R}^{3,1}$ which respects $SO(3)_q \times SO(1,1) \times {\bf Z}_2$ symmetry instead of the $ISO(2) \times SO(1,1) \times {\bf Z}_2$ symmetry of Bjorken flow.  (As before, the ${\bf Z}_2$ symmetry is the reflection $x^3 \to -x^3$, or equivalently $\eta \to -\eta$.)  I expect that symmetry constraints will completely determine $u^\mu$.  Invariance under ${\partial \over \partial\eta}$, ${\partial \over \partial\phi}$, and $\eta \to -\eta$ requires
 \eqn{uKappaForm}{
  u = \cosh\kappa(\tau,x_\perp) {\partial \over \partial\tau} + 
    \sinh\kappa(\tau,x_\perp) {\partial \over \partial x_\perp} \,.
 }
Readers familiar with the literature on hydrodynamical treatments of heavy-ion collisions will recognize \eno{uKappaForm} as a rewriting of the standard parametrization
 \eqn{StandardU}{
  u = \gamma_\perp \left( {\partial \over \partial\tau} + 
    v_x {\partial \over \partial x} + v_y {\partial \over \partial y} \right) \,,
 }
where
 \eqn{TransverseGamma}{
  \gamma_\perp = {1 \over \sqrt{1 - v_\perp^2}} \qquad\hbox{and}\qquad
    v_\perp = \sqrt{v_x^2 + v_y^2} \,.
 }
For clarity of comparison with the literature, in \eno{StandardU} and \eno{TransverseGamma} I have parametrized the transverse plane with $(x,y)$ instead of $(x^1,x^2)$.  The quantity $v_\perp$ is known as the transverse velocity.  Comparing \eno{uKappaForm} and \eno{StandardU} leads immediately to
 \eqn{ParameterRelations}{
  u^\tau = \cosh\kappa = \gamma_\perp \qquad
   {u^\perp \over u^\tau} = \tanh\kappa = v_\perp \,.
 }

Clearly, the vector $u$ in \eno{uKappaForm} commutes with ${\partial \over \partial\phi}$ and ${\partial \over \partial\eta}$: this is what ``invariance under ${\partial \over \partial\eta}$ and ${\partial \over \partial\phi}$'' means.  But there is no choice of $\kappa(\tau,x_\perp)$ that will get $u$ to commute with $\zeta$.  The closest one can come is to choose
 \eqn{kappaChoice}{
  \kappa = \arctanh {2q^2 \tau x_\perp \over 1 + q^2 \tau^2 + q^2 x_\perp^2} \,,
 }
in which case one finds
 \eqn{uSatisfies}{
  {\cal L}_\zeta u_\mu = {1 \over 4} (\nabla_\lambda \zeta^\lambda) u_\mu \,.
 }
The result \eno{uSatisfies} at first seems discouraging, but in fact it is ideal.  Let's refer to a tensor $Q_{\mu_1\mu_2\cdots}^{\nu_1\nu_2\cdots}$ as having $\zeta$-weight equal to $\alpha$ if
 \eqn{ZetaWeightDef}{
  {\cal L}_\zeta Q_{\mu_1\mu_2\cdots}^{\nu_1\nu_2\cdots} = 
    -{\alpha \over 4} (\nabla_\lambda \zeta^\lambda) Q_{\mu_1\mu_2\cdots}^{\nu_1\nu_2\cdots}
     \,.
 }
According to \eno{uSatisfies}, $u_\mu$ has $\zeta$-weight $-1$, and according to \eno{CKV}, the metric $g_{\mu\nu}$ has $\zeta$-weight $-2$.  Thus the projection tensor $g_{\mu\nu} + u_\mu u_\nu$ has $\zeta$-weight $-2$.  This projection tensor enters so ubiquitously into hydrodynamical equations that we should be delighted to see it transform as simply as possible under the symmetry generated by $\zeta$.  As I will explain in section~\ref{COVERING}, $\zeta$-weight is closely related to the more general notion of conformal weight.

\section{The inviscid case}
\label{INVISCID}

Having decided upon a four-velocity profile, the next step is to consider what the energy density $\epsilon$ should be.  Let's start by requiring that $\epsilon$ is invariant under ${\partial \over \partial\eta}$ and ${\partial \over \partial\phi}$: thus $\epsilon = \epsilon(\tau,x_\perp)$.  In the case $q=0$, demanding that also $\zeta^\mu \partial_\mu \epsilon = 0$ would lead immediately to the conclusion that $\epsilon$ depends only on $\tau$.  It is tempting to require $\zeta^\mu \partial_\mu \epsilon = 0$ even when $q \neq 0$, but the experience of finding \eno{uSatisfies} instead of the more obvious condition ${\cal L}_\zeta u_\mu = 0$ is a hint that a slightly more elastic notion of symmetry is appropriate.  Let us therefore consider the equation
 \eqn{zetaWeightEnergy}{
  {\cal L}_\zeta \epsilon = -{\alpha \over 4} (\nabla_\lambda \zeta^\lambda) \epsilon \,.
 }
Understanding $\epsilon$ as a scalar quantity, this equation can be read as saying that the function $\epsilon(\tau,x_\perp)$ has $\zeta$-weight $\alpha$.  It is easy to see that the general solution to \eno{zetaWeightEnergy} is
 \eqn{FoundEpsilonG}{
  \epsilon = {\hat\epsilon(g) \over \tau^\alpha} \qquad\hbox{where}\qquad
   g = {1 - q^2 \tau^2 + q^2 x_\perp^2 \over 2q\tau}
 }
and $\hat\epsilon(g)$ is an arbitrary function.  If one works to leading order in small $q$, one has $g \propto 1/\tau$, and \eno{FoundEpsilonG} reduces to the original conclusion that the energy density should depend only on $\tau$.

Some dynamical information is required to pin down what $\hat\epsilon(g)$ is.  The standard equations of viscous relativistic hydrodynamics are
 \eqn{Conservation}{
  \nabla^\mu T_{\mu\nu} = 0
 }
where
 \eqn{ViscousStress}{
  T_{\mu\nu} &\equiv \epsilon u_\mu u_\nu + p P_{\mu\nu} - 
    2 \eta \sigma_{\mu\nu} - \zeta (\nabla_\lambda u^\lambda) P_{\mu\nu}  \cr
  \sigma_{\mu\nu} &\equiv 
    P_\mu{}^\alpha P_\nu{}^\beta \left( {\nabla_\alpha u_\beta + \nabla_\beta u_\alpha
      \over 2} - 
      {g_{\alpha\beta} \over 3} \nabla_\lambda u^\lambda \right)  \cr
  P_{\mu\nu} &\equiv g_{\mu\nu} + u_\mu u_\nu \,.
 }
To complete the equations, we must specify how the pressure $p$, the shear viscosity $\eta$, and the bulk viscosity $\zeta$ depend on $\epsilon$.  Let's start with the simplest case:
 \eqn{ConformalInviscid}{
  p = {\epsilon \over 3} \qquad \eta = \zeta = 0 \,.
 }
The stress tensor is then traceless, as conformal invariance demands.  The conservation equations \eno{Conservation} overconstrain $\epsilon$ because both the $\nu=\tau$ and $\nu=x_\perp$ equations are non-trivial.  By inspection I found that these equations are consistent iff $\alpha=4$.\footnote{In retrospect it seems obvious that $\alpha=4$: then if $\hat\epsilon$ is dimensionless, \eno{FoundEpsilon} leads correctly to the conclusion that $\epsilon$ has dimension $4$.}  Then they reduce to
 \eqn{epsilonHatEq}{
  {d\log\hat\epsilon \over dg} = -{8g / 3 \over 1+g^2} \,,
 }
whose solution is
 \eqn{FoundEpsilonHat}{
  \hat\epsilon = {\hat\epsilon_0 \over (1+g^2)^{4/3}} \,.
 }
Using \eno{FoundEpsilonG}, one finds immediately
 \eqn{FoundEpsilon}{
  \epsilon = {\hat\epsilon_0 \over \tau^{4/3}} {(2q)^{8/3} \over 
   \left[ 1 + 2q^2 (\tau^2 + x_\perp^2) + 
    q^4 (\tau^2 - x_\perp^2)^2 \right]^{4/3}} \,,
 }
where $\hat\epsilon_0$ is an integration constant.  Recall that through \eno{uKappaForm} and \eno{kappaChoice}, we have already completely fixed $u^\mu$.  Also note that if we take $q \to 0$ with $\hat\epsilon_0 q^{8/3}$ held fixed, then from \eno{FoundEpsilon} we recover the standard result $\epsilon \propto 1/\tau^{4/3}$ for Bjorken flow.  Thus \eno{FoundEpsilon} together with \eno{uKappaForm} and \eno{kappaChoice} provide an exact solution of the equations of relativistic inviscid hydrodynamics with $p=\epsilon/3$ which becomes Bjorken flow in the limit $q \to 0$, but at finite $q$ describes a medium which has integrable falloff in the $x_\perp$ direction.  It is amusing to see an additional ${\bf Z}_2$ symmetry emerge which seems to have nothing to do with conformal symmetry: Based on \eno{uKappaForm}, \eno{kappaChoice}, and \eno{FoundEpsilon}, the quantities $\tau^{4/3} \epsilon$, $\kappa$, $u^\tau$, and $u^\perp$ are all invariant under the exchange of $\tau$ and $x_\perp$.

\section{The viscous case}
\label{VISCOUS}

The obvious thing to try next is to generalize to non-zero viscosity while preserving conformal invariance of the theory: that is,
 \eqn{ViscousHydro}{
  p = {\epsilon \over 3} \qquad \eta = {\rm H}_0 \epsilon^{3/4} \qquad
   \zeta = 0 \,.
 }
The $\epsilon^{3/4}$ dependence is necessary because only then is ${\rm H}_0$ dimensionless.  In searching for a solution with non-zero shear viscosity, I'm not going to change $u^\mu$ at all: recall that $u^\mu$ is fixed entirely by the requirement that $u^\mu$ should have $\zeta$-weight equal to $-1$.  I'm also not going to change the requirement that $\epsilon$ should have $\zeta$-weight equal to $4$: this is just the generalization to non-zero $q$ of the condition that $\epsilon$ should be constant across the transverse plane.  In other words, I'm going to assume \eno{uKappaForm}, \eno{kappaChoice}, and \eno{FoundEpsilon} with $\alpha=4$, and plug everything into \eno{ViscousHydro} and \eno{Conservation} to get equations for $\hat\epsilon(g)$.  Happily, the conservation equations are still consistent with one another for non-zero ${\rm H}_0$.  The equation they imply for $\hat\epsilon(g)$ is simpler to state in terms of
 \eqn{FourthRootEpsilon}{
  \hat{T}(g) = \sqrt[4]{\hat\epsilon(g)} \,,
 }
which is related to the local temperature of the fluid.  The conservation equations imply
 \eqn{ThatEquation}{
  3 (1+g^2)^{3/2} {d\hat{T} \over dg} + 2g \sqrt{1+g^2} \hat{T} + 
    g^2 {\rm H}_0 = 0 \,,
 }
whose general solution is
 \eqn{ThatSolution}{
  \hat{T}(g) = {\hat{T}_0 \over (1+g^2)^{1/3}} + 
    {{\rm H}_0 g \over \sqrt{1+g^2}} \left[ 1 - (1+g^2)^{1/6}
      {}_2F_1\left( {1 \over 2}, {1 \over 6}; {3 \over 2}; -g^2 \right)
     \right] \,,
 }
where $\hat{T}_0$ is an integration constant and ${}_2F_1$ denotes a hypergeometric function:
 \eqn{HypergeometricDef}{
  {}_2F_1(\alpha,\beta;\gamma;z) &\equiv 1 + {\alpha\beta \over c} z + 
   {\alpha(\alpha+1) \beta(\beta+1) \over \gamma(\gamma+1)} {z^2 \over 2}  \cr
   &\qquad{} + {\alpha(\alpha+1)(\alpha+2) \beta(\beta+1)(\beta+2) \over 
     \gamma(\gamma+1)(\gamma+2)} {z^3 \over 3!} + \ldots \,.
 }
There is a pathology in the solution \eno{ThatSolution}: $\hat{T}$ is negative for large enough $g$.  In fact, $\hat{T}$ has a single real root $g_*$, and $\hat{T} < 0$ when $g>g_*$.  In other words, the $\hat{T} < 0$ pathology arises when $\tau$ is too small and/or $x_\perp$ is too big.  It shouldn't dismay us unduly since a similar pathology already arises in the $q \to 0$ limit.  To see this, expand \eno{ThatSolution} at large $g$ (which is equivalent to small $q$) to find
 \eqn{TExpand}{
  {\hat{T}(g) \over \tau} = 
    \left[ \hat{T}_0 - {\rm H}_0 {\Gamma(1/2) \Gamma(-1/3) \over 2\Gamma(1/6)} 
     \right] {(2q)^{2/3} \over \tau^{1/3}} - {{\rm H}_0 \over 2\tau} + {\cal O}(g^{-2})
 }
Scaling $q \to 0$ and $\hat{T}_0 \to \infty$ so that $\hat{T}_0 q^{2/3}$ remains finite, one finds
 \eqn{BjorkenViscous}{
  \sqrt[4]{\epsilon(\tau)} = {\hat{T} \over \tau} = 
    {e_0 \over \tau^{1/3}} - {{\rm H}_0 \over 2\tau} \,.
 }
It can be checked that all the corrections to \eno{BjorkenViscous} from the terms labeled ${\cal O}(g^{-2})$ in \eno{TExpand} come with positive powers of $q$ after the scaling just mentioned.  Therefore, \eno{BjorkenViscous} combined with the $q \to 0$ limit of the four-velocity profile, namely $u = \partial_\tau$, forms an exact solution to the equations of conformal viscous hydrodynamics.  This is viscous Bjorken flow.  The pathology at $\tau=\tau_*$, where
 \eqn{tauStarDef}{
  \tau_* = \left( {{\rm H}_0 \over 2e_0} \right)^{3/2} \,,
 }
can be understood as an indication that as $\tau$ approaches $\tau_*$ from above, eventually hydrodynamics cannot be used, because the shear is so strong that the viscous correction is more important than the pressure.  In such a situation, it must be expected that higher derivative corrections also become important.  The difficulties one anticipates as $g$ approaches $g_*$ from below in the finite $q$ case are essentially the same as when $\tau$ approaches $\tau_*$ from above for $q=0$.

\section{Before the collision}
\label{EARLY}

I'd now like to inquire what kind of non-hydrodynamical initial state might lead, at least in some approximation, to the hydrodynamical flow that I explained in the previous subsections.  A natural ansatz in heavy-ion collisions is to assume that the state before the collision can be described as sum of left- and right-moving parts, each of which moves at the speed of light:
 \eqn{ShockSum}{
  T_{uu} = T_{uu}(u,x_\perp) \qquad
  T_{vv} = T_{vv}(v,x_\perp) \,,
 }
where
 \eqn{uvDefs}{
  u = t - x^3 \qquad v = t + x^3 \,.
 }
All components of the stress tensor other than $T_{uu}$ and $T_{vv}$ are assumed to be zero in the coordinate system $(u,v,x_\perp,\phi)$ on ${\bf R}^{3,1}$.  

In order for the initial state \eno{ShockSum} to lead approximately to an $SO(3)_q \times SO(1,1) \times {\bf Z}_2$ invariant hydrodynamical flow, it would help for the initial state to preserve as much of this symmetry group as possible.  The $SO(1,1)$ boost symmetry can't possibly be preserved, because boosting a lightlike collision changes the total energy of the two participants multiplicatively while preserving the product of their energies.  $SO(1,1)$ symmetry is supposed to emerge in the mid-rapidity region through some post-collision dynamics.  But $SO(3)_q$ symmetry, as the analog of $ISO(2)$ symmetry in the transverse plane, is something we might sensibly demand of the initial state in a perfectly central collision.  Let's inquire what constraints on $T_{uu}$ and $T_{vv}$ arise when we do require $SO(3)_q$ symmetry.

The ansatz \eno{ShockSum} is trivially invariant under $\phi$ rotations, so the only issue is how to implement symmetry under the conformal Killing vector $\zeta$, which in $(u,v,x_\perp,\phi)$ coordinates reads
 \eqn{ZetaUV}{
  \zeta = 2 q^2 x_\perp \cos\phi 
   \left( u {\partial \over \partial u} + v {\partial \over \partial v} \right) + 
    (1 + q^2 uv + q^2 x_\perp^2) \cos\phi {\partial \over \partial x_\perp} - 
    {1 + q^2 uv - q^2 x_\perp^2 \over x_\perp} \sin\phi {\partial \over \partial\phi}
      \,.
 }
We have learned from previous sections that the useful notion of symmetry under $\zeta$ involves $\zeta$-weights.  When $\epsilon$ has $\zeta$-weight $4$, $u_\mu$ has $\zeta$-weight $-1$, and $g_{\mu\nu}$ has $\zeta$-weight $-2$, the hydrodynamical stress tensor $T_{\mu\nu}$ has $\zeta$-weight $2$.  This is not trivial to verify when the shear viscosity is non-zero, but it is true.\footnote{The Weyl covariant derivative introduced in \cite{Loganayagam:2008is} makes it substantially easier to check that $\sigma_{\mu\nu}$ has $\zeta$-weight $-1$.}  The only sensible way to respect symmetry under $\zeta$ prior to the collision is to demand that $T_{\mu\nu}$ should again have $\zeta$-weight $2$: that is,
 \eqn{LTexpress}{
  {\cal L}_\zeta T_{\mu\nu} = -{1 \over 2} (\nabla_\lambda \zeta^\lambda) T_{\mu\nu} \,.
 }
In short, the problem to be solved in this section is to find a solution to \eno{LTexpress} of the form \eno{ShockSum}.  For simplicity I'll also demand that the initial state preserve the ${\bf Z}_2$ symmetry, which exchanges $u$ and $v$: this just means that the left- and right-moving participants are identical.  The answer can be anticipated from AdS/CFT: directly from \cite{Gubser:2008pc} we can read off the result
 \eqn{FoundF}{
  T_{uu} = {2q^2 E \over \pi (1+q^2 x_\perp^2)^3} \delta(u) \qquad
  T_{vv} = {2q^2 E \over \pi (1+q^2 x_\perp^2)^3} \delta(v) \,.
 }
Here $E$ is the energy in one of the shocks.

Let me now show how \eno{FoundF} follows directly from \eno{LTexpress} together with the ansatz \eno{ShockSum}.  In $(u,v,x_\perp,\phi)$ coordinates,
The $\mu=u$, $\nu=x_\perp$ component of \eno{LTexpress} vanishes only if $u T_{uu} = 0$.  So $T_{uu} = e(x_\perp) \delta(u)$ for some function $e(x_\perp)$.  Plugging this expression into the $\mu=u$, $\nu=u$ component of \eno{LTexpress} leads immediately to
 \eqn{DistributionalEq}{
  (1 + q^2 x_\perp^2) e'(x_\perp) \delta(u) + 2 q^2 x_\perp e(x_\perp)
    (4\delta(u) + u \delta'(u)) = 0 \,.
 }
This can be simplified by using the distributional identity $u \delta'(u) = -\delta(u)$: then \eno{DistributionalEq} becomes
 \eqn{DistSimplified}{
  (1+q^2 x_\perp^2) e'(x_\perp) + 6 q^2 x_\perp e(x_\perp) = 0 \,,
 }
which can easily be solved to find the expression for $T_{uu}$ given in \eno{FoundF}.  A similar argument applies to $T_{vv}$, based on the $vx_\perp$ and $vv$ components of \eno{LTexpress}.  It is straightforward to check that all other components of the equation \eno{LTexpress} are satisfied.

Already in \cite{Gubser:2008pc} it was remarked that the transverse profile of $T_{uu}$ in \eno{FoundF} differs significantly from a highly boosted Woods-Saxon profile: See in particular figure~4 of that work.  If $q$ is chosen so that the energy-weighted root-mean-square (rms) value of $x_\perp$ is the same between the $SO(3)_q$-symmetric and Woods-Saxon profiles, then the principal difference between the two is that there is more weight near $x_\perp=0$ in the $SO(3)_q$-symmetric profile.  It is also significant that the large $x_\perp$ tail of the $SO(3)_q$-symmetric profile is qualitatively larger than for Woods-Saxon: power-law falloff as compared to exponential.  Ideally, one should develop a theory of how deviations from $SO(3)_q$ symmetry in the initial state propagate to the hydrodynamical stage of the collision.

\section{Plugging in approximate numbers}
\label{NUMBERS}

Because my motivation was to understand the hydrodynamic phase of the quark-gluon plasma produced in heavy-ion collisions, I would now like to plug in numbers which are at least approximately representative of a real-world gold-gold collision at top RHIC energies.  For some of the simpler quantities I will follow \cite{Gubser:2008pc}:
 \eqn{SimplerQuantities}{
  {1 \over q} &= 4.3\,{\rm fm} \qquad
  E = E_{\rm beam} = 19.7\,{\rm TeV} \qquad
  f_* \equiv {\epsilon \over T^4} = 11 \,.
 }
If the inviscid flow \eno{FoundEpsilon} is our goal, then it remains only to provide a value for the dimensionless quantity $\hat\epsilon_0$.  If we want to discuss the viscous solution \eno{ThatSolution} quantitatively, then we must instead provide values for the dimensionless quantities $\hat{T}_0$ and ${\rm H}_0$.  

An obvious plan for getting at $\hat\epsilon_0$ or $\hat{T}_0$ is to compute the entropy per unit rapidity in the fluid and then compare with phenomenological estimates of the same quantity.\footnote{In the interests of a compact presentation, I will not distinguish between rapidity and pseudo-rapidity.}  In the final, hadronized state, entropy is related to the number of charged tracks:
 \eqn{ChargedTracks}{
  {dS \over d\eta} \approx 7.5 {dN_{\rm charged} \over d\eta} \approx 5000 \,.
 }
The reader interested in details of where \eno{ChargedTracks} comes from is again referred to \cite{Gubser:2008pc} and references therein, particularly \cite{Pal:2003rz}.  $dN_{\rm charged} / d\eta$ is directly measurable, and for the most central collisions at $\sqrt{s_{\rm NN}} = 200\,{\rm GeV}$, a reasonable figure is $dN_{\rm charged} / d\eta = 660$ \cite{Back:2004je}.  This is the figure that went into the second approximate equality in \eno{ChargedTracks}.\footnote{Because \eno{ChargedTracks} refers to final state entropy, the entropy of the fluid per unit rapidity might be significantly lower.  Indeed, using the estimates of energy density in \cite{Adcox:2004mh} one finds $(dS/d\eta)_{\rm fluid} \approx 3000$: see appendix A.  To be more systematic in my treatment of $dS/d\eta$, I would have to introduce some definite assumptions about hadronization.  This would take me too far afield from my main purpose, which is to obtain approximate numbers for the hydrodynamical flows I have found.}

The entropy density can be determined from the energy density:
 \eqn{EntropyDensity}{
  s = \Sigma_0 \epsilon^{3/4} \qquad\hbox{where}\qquad
   \Sigma_0 = {4 \over 3} f_*^{1/4} = 2.43 \,.
 }
In order to compute $dS/d\eta$ for the hydrodynamical flows found in previous sections, we must use the entropy current,
 \eqn{EntropyCurrent}{
  s^\mu = s u^\mu \,,
 }
where $s$ is given by \eno{EntropyDensity}.  If $M$ is a co-dimension $1$ surface, with coordinates $y^\alpha$, whose induced metric is $h_{\alpha\beta}$ and whose unit normal vector $n^\mu$ is everywhere timelike, then the entropy on $M$ is
 \eqn{SMDef}{
  S_M = \int_M d^3 y \, \sqrt{\det h_{\alpha\beta}} \, n^\mu s_\mu \,.
 }
If we take $M$ to be a slice of constant $\tau$, and use coordinates $y^\alpha = (\eta,x_\perp,\phi)$, then the integral in \eno{SMDef} diverges because the integrand is $\eta$-independent---due precisely to the boost symmetry.  But we can define the entropy per unit rapidity as
 \eqn{dSdeta}{
  {dS \over d\eta} = 2\pi \tau \int_0^\infty x_\perp dx_\perp \, s^\tau
    = 2\pi \Sigma_0 \tau \int_0^\infty x_\perp dx_\perp \, \epsilon^{3/4} u^\tau \,.
 }
For the inviscid flow, where we take $\epsilon$ from \eno{FoundEpsilon} and $u^\tau$ from \eno{uKappaForm} and \eno{kappaChoice}, the integral in \eno{dSdeta} can be done explicitly, and one finds
 \eqn{InviscidS}{
  {dS \over d\eta} = 4\pi \Sigma_0 \hat\epsilon_0^{3/4} \,.
 }
There is no $\tau$ dependence because entropy is conserved by inviscid flows.  There is also no dependence on $q$ in \eno{InviscidS}, which could be anticipated since $q$ is dimensionful and none of the other quantities is.  Comparing \eno{ChargedTracks} and \eno{InviscidS}, and using \eno{EntropyDensity}, one finds
 \eqn{EpsilonNumber}{
  \hat\epsilon_0 = {1 \over f_*^{1/3}} 
    \left( {3 \over 16\pi} {dS \over d\eta} \right)^{4/3} \approx 880 \,.
 }

\begin{figure}[t]
  \centerline{\includegraphics[width=7in]{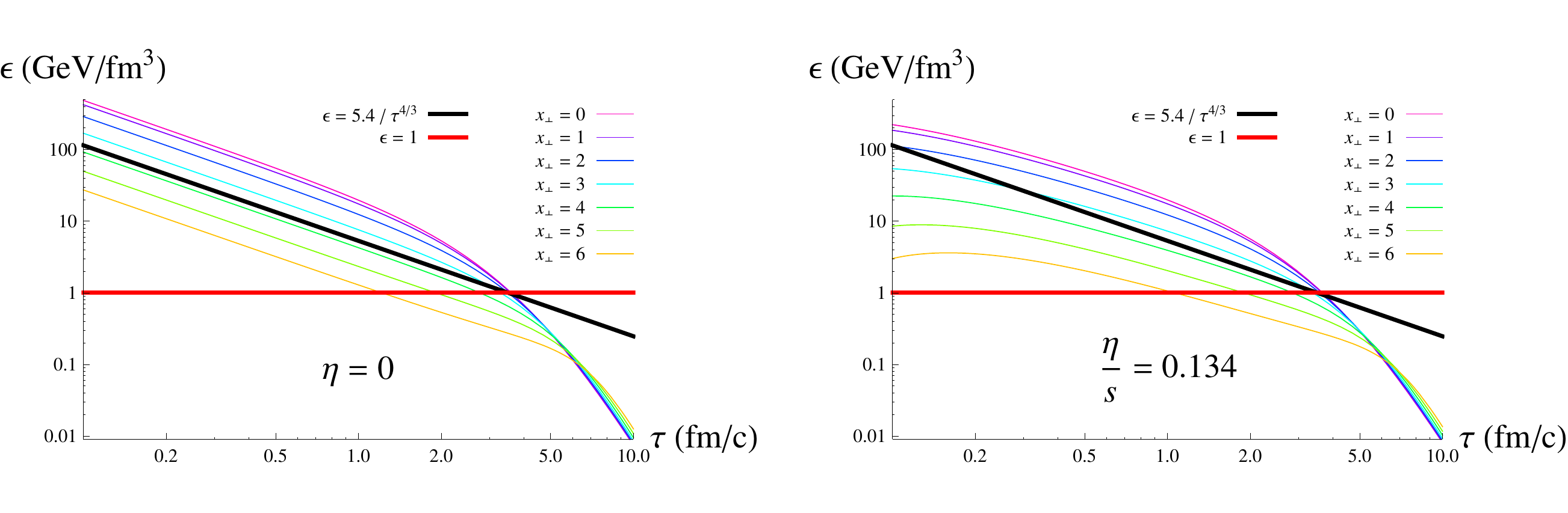}}
  \caption{Left: $\epsilon$ as a function of $\tau$ for different values of $x_\perp$ at zero viscosity, with parameters chosen as in \eno{SimplerQuantities} and \eno{EpsilonNumber}.  Right: $\epsilon$ for non-zero viscosity, with parameters chosen as in \eno{SimplerQuantities}, \eno{GuessEta}, and \eno{GuessTo}.  The bold red line shows the dependence $\epsilon = 5.4/\tau^3$, where $\epsilon$ is in ${\rm GeV}/{\rm fm}^3$ and $\tau$ is in ${\rm fm}/c$.  The estimate $\epsilon = 5.4\,{\rm GeV}/{\rm fm}^3$ at $\tau = 1\,{\rm fm}/c$ is taken from \cite{Adcox:2004mh}.}\label{epsPlots}
 \end{figure}

\begin{figure}[t]
  \centerline{\includegraphics[width=5in]{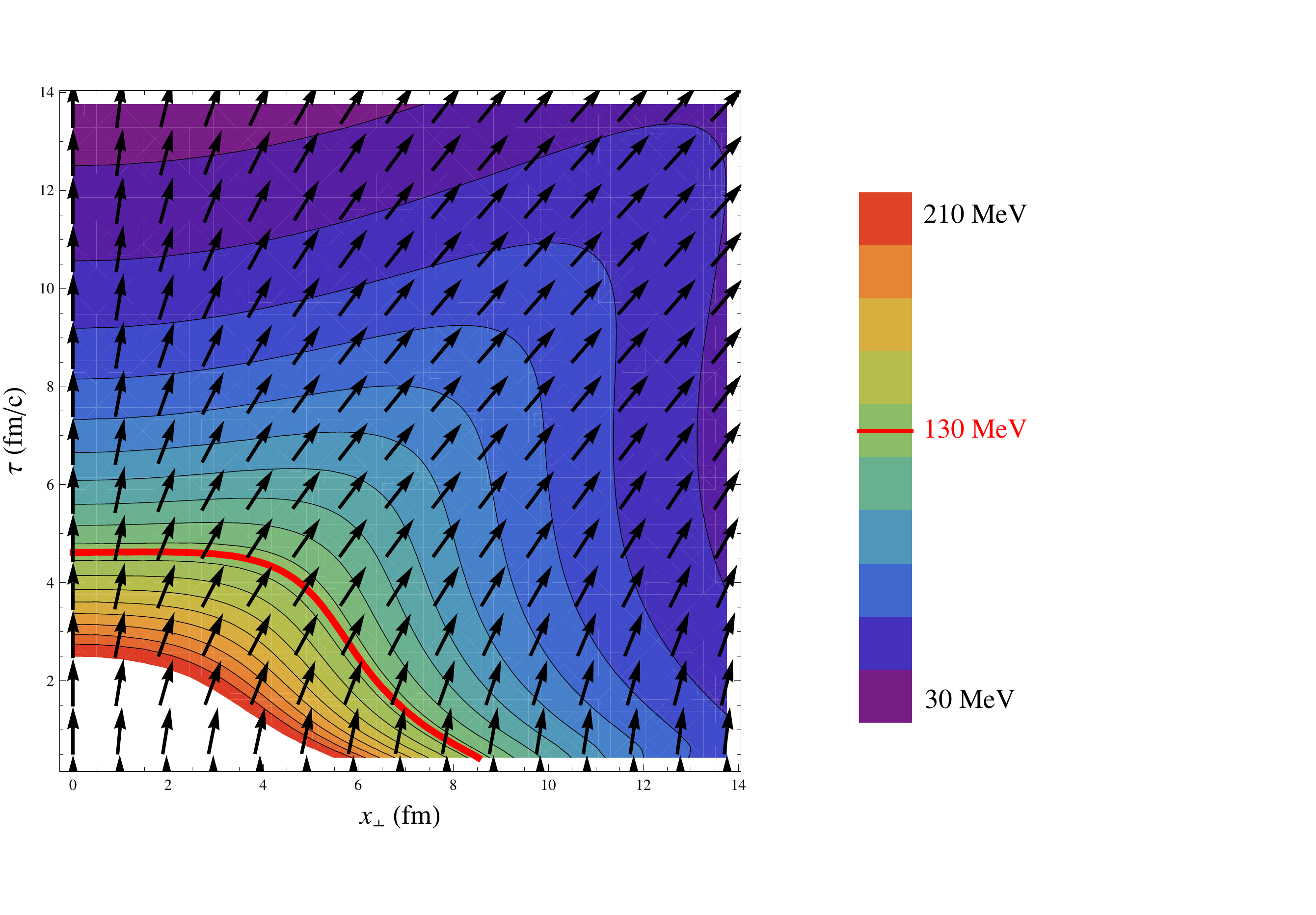}}
  \caption{The vector field $u^\mu$ and the temperature profile for the viscous flow, with parameters chosen as in \eno{SimplerQuantities}, \eno{GuessEta}, and \eno{GuessTo}.  To improve readability I have plotted not $(u^\tau,u^\perp)$ but instead $(u^\tau,u^\perp)/\sqrt{(u^\tau)^2+(u^\perp)^2}$.  The thick red contour is $T = 130\,{\rm MeV}$.  The cooler parts of the plot have little to do with heavy-ion phenomenology, but they help illustrate the nature of the flow.}\label{FlowShape}
 \end{figure}

\begin{figure}[t]
  \centerline{\includegraphics[width=7in]{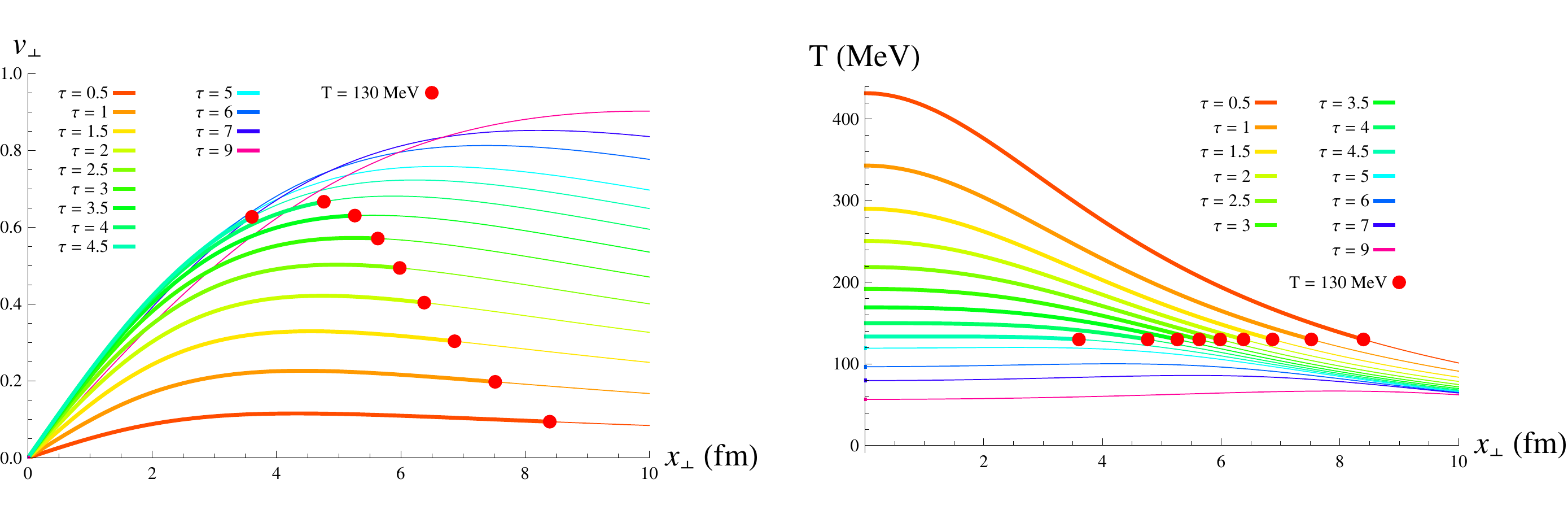}}
  \caption{Left: The transverse velocity $v_\perp = \tanh\kappa$ as a function of $x_\perp$ for selected values of $\tau$, measured in ${\rm fm}/c$, with $1/q = 4.3\,{\rm fm}$ as in \eno{SimplerQuantities}.  Right: The local temperature $T$ as a function of $x_\perp$ for selected values of $\tau$ with non-zero viscosity, with parameters chosen as in \eno{SimplerQuantities}, \eno{GuessEta}, and \eno{GuessTo}.  In both plots, the red dots show where $T = 130\,{\rm MeV}$.  For $\tau$ \raise2.5pt\hbox{$>$}\hskip-0.12in\lower2.5pt\hbox{$\sim$} $4.6\,{\rm fm}/c$, 
the temperature is below $130\,{\rm MeV}$ everywhere.}\label{vTPlots}
 \end{figure}

To treat viscous flow, we must decide on a value for the shear viscosity.  I will take as a representative number the lattice result $\eta/s = 0.134$ for $SU(3)$ gluodynamics \cite{Meyer:2007ic}: a bit larger than the value $\eta/s = 1/4\pi$ found in \cite{Policastro:2001yc,Kovtun:2004de}.  Thus
 \eqn{GuessEta}{
  {\rm H}_0 = {\eta \over s} \Sigma_0 = 0.33 \,.
 }
Non-zero shear viscosity implies that entropy increases with time.  Because \eno{ChargedTracks} refers to final state entropy, we should compare it with $dS/d\eta$ on a fairly late time-slice.  I will again evaluate entropy on a surface $M$ at constant $\tau$.  Recalling that the temperature formally becomes negative at large enough $x_\perp$, I see that I have to cut $M$ off at some limiting value $x_{\perp*}$.  To find $x_{\perp*}$, one must solve the equation
 \eqn{ToFindXperp}{
  g_* = {1 - q^2 \tau^2 + q^2 x_{\perp*}^2 \over 2q\tau}
 }
for $x_{\perp*}$, where $g_*$ is the unique real root of the equation $\hat{T}(g) = 0$.  In short,
 \eqn{dSdetaViscous}{
  {dS \over d\eta} = 2\pi\tau \int_0^{x_{\perp*}} x_\perp dx_\perp \, s^\tau \,,
 }
where $s^\tau$ is computed as before, only using the full viscous solution \eno{ThatSolution}.  In order to have \eno{dSdetaViscous} match the result \eno{ChargedTracks} with ${\rm H}_0 = 0.33$, one needs
 \eqn{GuessTo}{
  \hat{T}_0 = 5.55 \,.
 }
This is to be compared to $\sqrt[4]{\hat\epsilon_0} = 5.45$ from the inviscid flow based on \eno{EpsilonNumber}.  In figure~\ref{epsPlots} I show the time evolution of $\epsilon(\tau,x_\perp)$ for several values of $x_\perp$.  For energy density I have used the standard units ${\rm GeV}/{\rm fm}^3$.  To convert to units of ${\rm fm}^{-4}$, one need only recall that $\hbar c = 0.197327 \, {\rm GeV} \, {\rm fm}$ and set $\hbar=c=1$.  In figure~\ref{FlowShape} I show the direction of the hydrodynamic flow and contours of constant temperature for the viscous flow.  In figure~\ref{vTPlots} I show the dependence of transverse velocity $v_\perp = \tanh\kappa$ and the temperature $T$ on $x_\perp$ for several fixed values of $\tau$, using the viscous flow.

Phenomenologically oriented readers will notice in these last plots undesirably large $v_\perp$ and $T$ for large $x_\perp$.  These tails are a consequence of assuming {\it exact} conformal symmetry in the underlying state and of supplying initial states whose stress tensor has only power-law fall-off at large $x_\perp$.  To understand the tails better, let's consider what they look like at fixed $\tau > 0$.  From \eno{kappaChoice} one sees immediately that $v_\perp \propto 1/x_\perp$ for large $x_\perp$.  From \eno{ThatSolution} it follows that the temperature falls to $0$ at the limiting value $x_{\perp*}$ discussed around \eno{dSdetaViscous}.  This strange behavior is associated with a complete breakdown of the derivative expansion on which hydrodynamics is based.

In more practical terms, the region of the flows where $T < 130\,{\rm MeV}$ should probably be regarded as phenomenologically unuseful: $130\,{\rm MeV}$ is approximately the temperature of decoupling in a Cooper-Frye treatment, and it is significantly below $T_c \approx 170\,{\rm MeV}$, so neither the conformal approximation nor the hydrodynamic approximation is any good when $T < 130\,{\rm MeV}$.  In short, one could regard the surface $T = 130\,{\rm MeV}$ as a freeze-out surface beyond which the true degrees of freedom are nearly free hadrons rather than a locally equilibrated fluid.  It is worth noting that at early times, most of the energy is in the region where $T>130\,{\rm MeV}$, even though the region $x_\perp < x_{\perp*}$ is substantially larger.  In particular, for the viscous flow at $\tau = 0.5\,{\rm fm}/c$, slightly more than $97\%$ of the energy is inside the freeze-out surface, which is at $x_\perp = 8.4\,{\rm fm}$, whereas $T$ (as defined from \eno{ThatSolution}) drops to zero at $x_{\perp*} = 29.5\,{\rm fm}$.

\begin{figure}
  \centerline{\includegraphics[width=4in]{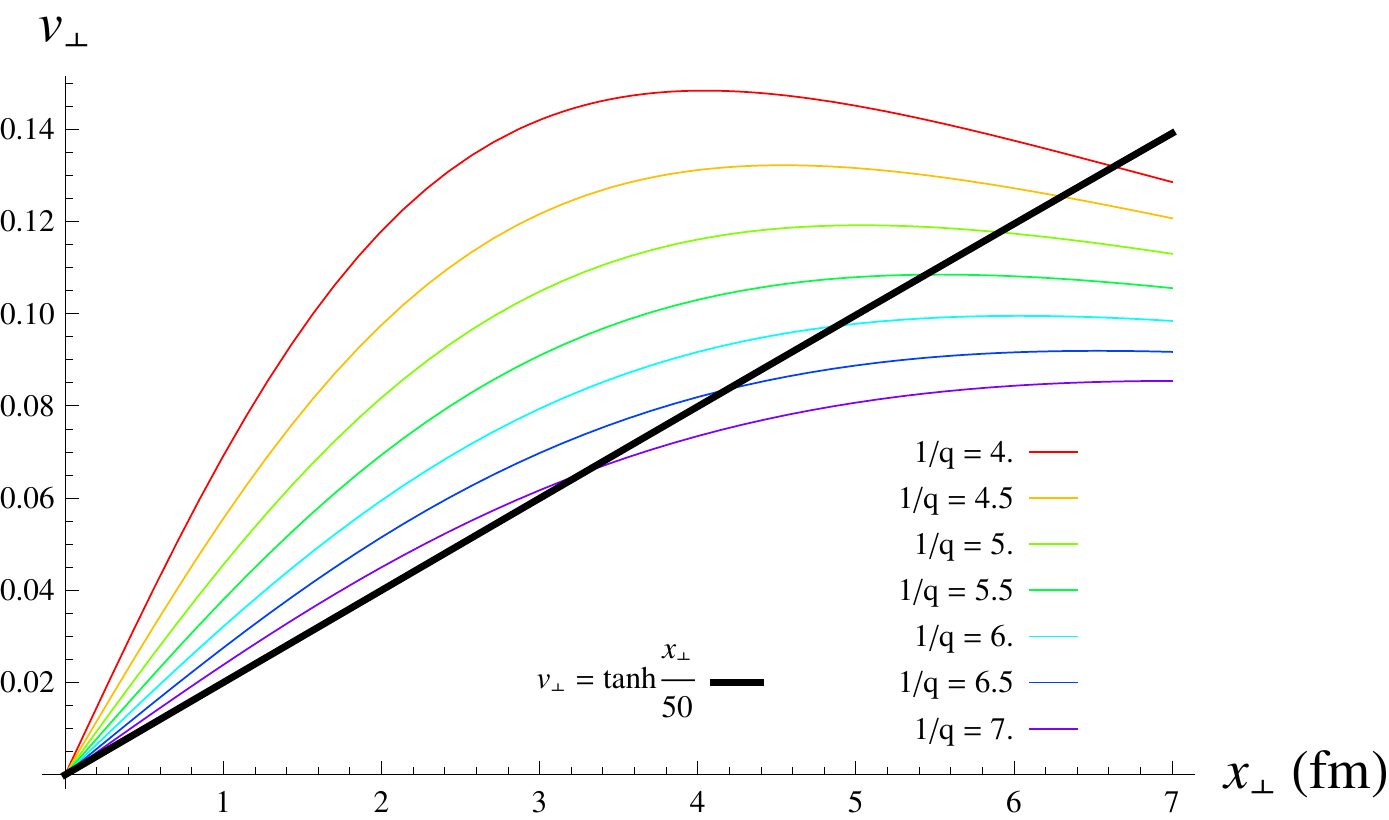}}
  \caption{The transverse velocity $v_\perp$ as a function of transverse radius $x_\perp$ at $\tau = 0.6\,{\rm fm}/c$ for several values of $1/q$, measured in ${\rm fm}$.  The dark line is the phenomenological proposal $v_\perp = \tanh {x_\perp \over 50}$ of \cite{Kolb:2002ve}.}\label{SeveralQ}
 \end{figure}

The value of $q$ in \eno{SimplerQuantities} is the one which makes the energy-weighted rms transverse radius match between the $SO(3)_q$-invariant stress tensor profile \eno{FoundF} and the highly boosted Woods-Saxon profile of a gold nucleus.  In other words, I chose the $SO(3)_q$ symmetry to be the one most nearly realized by the pre-collision state, and then I used that symmetry (assumed to be exact and to be preserved by the underlying dynamics) to constrain $u^\mu$ and $T_{\mu\nu}$ after the collision.  While this idea seems suitably straightforward for a first attempt, it could be in need of refinement to account for effects of evolution of the parton distribution function, initial state radiation, and other early time dynamics.  As \eno{InviscidS} makes clear, $\hat\epsilon_0$ for the inviscid flow is entirely independent of $q$.  Also, ${\rm H}_0$ is independent of $q$ because it amounted simply to a choice of $\eta/s$.  Finally, $\hat{T}_0$ is only weakly dependent on $q$ when it is determined by matching \eno{dSdetaViscous} with the observed multiplicity per unit rapidity.  Thus we can think of varying $q$ independently of these other parameters.  An interesting quantity which depends strongly on $q$ is the transverse velocity
 \eqn{TransverseV}{
  v_\perp = \tanh\kappa = {2q^2 \tau x_\perp \over 1 + q^2 \tau^2 + q^2 x_\perp^2} \,.
 }
In figure~\ref{SeveralQ} I plot $v_\perp$ at $\tau=0.6\,{\rm fm}$ for several different values of $q$ and show for comparison a radial flow profile studied in \cite{Kolb:2002ve}.  The comparison seems to favor $1/q$ somewhat larger than $4.3\,{\rm fm}$.  Energy density, entropy density, and temperature get smaller when $1/q$ is made larger, simply because we are expanding the flow in all space-time dimensions without increasing total entropy.  However, the freeze-out surface changes only slowly with $1/q$.  It would be interesting to work through a more systematic study of single-particle spectra and Hanbury Brown-Twiss radii with \eno{TransverseV} as an initial condition for the hydrodynamic flow: then one could establish a preferred value of $q$ purely from final-state observables.

\section{Mapping to the covering space}
\label{COVERING}

Exact solutions to the Navier-Stokes equations are few and far between.  I was able to get hold of the one described in this paper because I imposed enough symmetry constraints so that there was effectively only one independent variable left, namely $g$ defined in \eno{FoundEpsilon}.  This variable is essentially the only combination of the coordinates on ${\bf R}^{3,1}$ which is invariant under $SO(3)_q \times SO(1,1)$.  It is instructive to understand what $g$ looks like when we conformally embed Minkowski space in its covering space $S^3 \times {\bf R}$.  An explicit mapping to accomplish this can be found, for example, in \cite{Friess:2006kw}, and can be characterized as follows.  Parametrize ${\bf R}^{3,1}$ by $(t,r,\theta,\phi)$, where $t$ is the usual lab time, $r = \sqrt{(x^1)^2 + (x^2)^2 + (x^3)^2}$, and $(\theta,\phi)$ are standard coordinates on $S^2$.  Parametrize $S^3 \times {\bf R}$ by $(\sigma,\chi,\theta,\phi)$, where $\sigma$ is global time (for the ${\bf R}$ piece), $(\theta,\phi)$ are coordinates on $S^2$ as before, and $\chi$ is an angle describing the dimension orthogonal to the $S^2$ parametrized by $(\theta,\phi)$.  Let's make all quantities on $S^3 \times {\bf R}$ dimensionless, so that the metric is
 \eqn{CSmetric}{
  ds^2 = -d\sigma^2 + d\chi^2 + \sin^2 \chi \, (d\theta^2 + \sin^2 \theta \, d\phi^2) \,.
 }
Then the embedding of ${\bf R}^{3,1}$ into $S^3 \times {\bf R}$ is specified by
 \eqn{SpecifyMap}{
  qt = W \sin\sigma \qquad qr = W \sin\chi
 }
where
 \eqn{Wdef}{
  W = {1 \over \cos\sigma + \cos\chi} \,.
 }
It appears from \eno{SpecifyMap} and \eno{Wdef} that $\sigma$ is an angular coordinate, to be identified modulo $2\pi$.  In fact, one period of $\sigma$ is enough to cover ${\bf R}^{3,1}$.

One can obviously embed $S^3$ in ${\bf R}^4$ as
 \eqn{yEmbed}{
  \begin{pmatrix} y_1 \\ y_2 \\ y_3 \\ y_4 \end{pmatrix} = 
    \begin{pmatrix} \sin\chi \sin\theta \cos\phi \\
     \sin\chi \sin\theta \sin\phi \\
     \sin\chi \cos\theta \\ \cos\chi \end{pmatrix} \,.
 }
What I have called $SO(3)_q$ is simply the $SO(3)$ which rotates $y_1$, $y_2$, and $y_4$ among themselves while leaving $y_3$ fixed.  Slightly tedious manipulations starting from \eno{SpecifyMap} and \eno{Wdef}, together with $x^3 = r \cos\theta$, lead to
 \eqn{gExpressions}{
  g \equiv {1 + q^2 x_\perp^2 - q^2 \tau^2 \over 2 q \tau} = 
    {\cos\sigma \over \sqrt{\sin^2 \sigma - y_3^2}} \,.
 }
The last expression is obviously $SO(3)$-invariant, since $\sigma$ and $y_3$ are separately.  But $SO(1,1)$-invariance, which is obvious from the middle expression in \eno{gExpressions}, is far from obvious in the coordinates on $S^3 \times {\bf R}$.  When $r=0$, then also $x_\perp = y_3=0$, and $g = \cot\sigma$ is a function simply of global time.  Away from $r=0$, $g$ depends also on the ``latitude'' variable $y_3$ on $S^3$.  Note that slices of constant $y_3$ on $S^3$ are two-spheres, though not the ones parametrized by $(\theta,\phi)$.  The $SO(3)$ symmetry acts to rotate these two-spheres.

To further understand the geometry of the flow, let's start with a general description of conformal mappings and conformal weights.  Given two manifolds with metrics, $(M,g_{\mu\nu})$ and $(\tilde{M},\tilde{g}_{\mu\nu})$, a conformal map from $M$ to $\tilde{M}$ is a smooth bijection $x^\mu \to \tilde{x}^\mu$ such that
 \eqn{MetricMap}{
  \tilde{g}_{\mu\nu} = {1 \over W^2} {\partial x^\kappa \over \partial\tilde{x}^\mu}
    {\partial x^\lambda \over \partial\tilde{x}^\nu} g_{\kappa\lambda} \,,
 }
where $W$ is the conformal factor.  If $W=1$, then the map is an isometry.  For the map \eno{SpecifyMap}, the conformal factor is given in \eno{Wdef}.  A tensor $Q_{\mu_1\mu_2\cdots}^{\nu_1\nu_2\cdots}$ on $M$ maps to a new tensor $\tilde{Q}_{\mu_1\mu_2\cdots}^{\nu_1\nu_2\cdots}$ on $\tilde{M}$ with conformal weight $\alpha$ iff
 \eqn{Qmaps}{
  \tilde{Q}_{\mu_1\mu_2\cdots}^{\nu_1\nu_2\cdots} = W^\alpha
    {\partial x^{\kappa_1} \over \partial\tilde{x}^{\mu_1}}
    {\partial x^{\kappa_2} \over \partial\tilde{x}^{\mu_2}} \cdots
    {\partial\tilde{x}^{\nu_1} \over \partial x^{\lambda_1}}
    {\partial\tilde{x}^{\nu_2} \over \partial x^{\lambda_2}} \cdots
    Q_{\kappa_1\kappa_2\cdots}^{\lambda_1\lambda_2\cdots} \,.
 }
My previous notion of $\zeta$-weight is in a sense a special case of \eno{Qmaps}: $Q_{\mu_1\mu_2\cdots}^{\nu_1\nu_2\cdots}$ on ${\bf R}^{3,1}$ has $\zeta$-weight $\alpha$ iff it maps to itself with conformal weight $\alpha$ upon the conformal map $x^\mu \to x^\mu + \vartheta \zeta^\mu$, where $\vartheta$ is an infinitesimal parameter which formally squares to $0$.

The four-velocity vector $u^\mu$ is naturally a tensor of conformal weight $-1$, because then its image $\tilde{u}^\mu$ is a timelike unit vector with respect to the new metric $\tilde{g}_{\mu\nu}$.  One can check that for the conformal map ${\bf R}^{3,1} \to S^3 \times {\bf R}$ discussed above, the only non-zero components of $\tilde{u}^\mu$ are in the $\sigma$ direction and the ``latitude'' direction, orthogonal to surfaces of constant $y_3$.\footnote{One can go further and show that if a tensor on ${\bf R}^{3,1}$ respects the $SO(3)_q$ symmetry with $\zeta$-weight $\alpha$, then mapping it with conformal weight $\alpha$ to $S^3 \times {\bf R}$ with conformal weight $\alpha$ results in a tensor on $S^3 \times {\bf R}$ which has vanishing Lie derivative under the Killing vectors that generate the $SO(3)$ that preserves surfaces of constant $y_3$.  A special case of this general claim is that the energy density on $S^3$ (ignoring the conformal anomaly of the stress tensor) is $\tilde\epsilon = W^4 \epsilon = {W^4 \over \tau^4} \hat\epsilon$, which is $SO(3)$-invariant because both $\hat\epsilon$ and ${W \over \tau} = q (\sin^2 \sigma - y_3^2)^{-1/2}$ are $SO(3)$-invariant.}  One can also check that the initial shock wave states discussed in section~\ref{EARLY} get mapped to shocks whose energy density is uniform across two-spheres of fixed $y_3$ and which reach $y_3=0$ at global time $\sigma = 0$.  Thus the overall picture on $S^3 \times {\bf R}$ is as much like the original Bjorken picture as it can be: uniform light-like shocks collide, and a boost-invariant fluid results which is uniform in the transverse directions.

\section{Summary}
\label{SUMMARY}

In order to accommodate finite transverse size and non-zero radial velocity in collisions of heavy ions, I propose to replace translation invariance in the transverse plane by symmetry under special conformal transformations, one of which is
 \eqn{ZetaOnceMore}{
  \zeta = {\partial \over \partial x^1} + q^2 \left[ 2x^1 x^\mu 
    {\partial \over \partial x^\mu} - x^\mu x_\mu {\partial \over \partial x^1} \right] \,,
 }
where $q$ is a parameter with dimensions of inverse length.  Along with rotations around the beamline, these special conformal transformations fill out an $SO(3)$ subgroup of the conformal group $SO(4,2)$, which is an approximate symmetry of QCD at high energies.  I denote this $SO(3)$ subgroup $SO(3)_q$.  A shock-wave traveling at the speed of light in the $+x^3$ direction which respects $SO(3)_q$ must take the form
 \eqn{TuuForm}{
  T_{uu} = {2q^2 E \over \pi (1 + q^2 x_\perp^2)^3} \delta(u) \,,
 }
where $E$ is the energy of the shock wave.  Comparing with a boosted nucleus whose energy density is assumed to follow the Woods-Saxon profile that describes gold nuclei, one finds $1/q = 4.3\,{\rm fm}$ in order to get the same energy-weighted root-mean-square transverse radius.  The form \eno{TuuForm} can be derived straightforwardly from the gauge-string duality as the dual of a point-sourced shock wave in $AdS_5$ \cite{Gubser:2008pc}; however, like all the results in this paper, it doesn't depend at all on the dynamical content of the gauge-string duality.

The only four-velocity profile which respects $SO(3)_q$ in addition to $SO(1,1)$ boost invariance along the beamline and the ${\bf Z}_2$ symmetry that reflects $x^3 \to -x^3$ is
 \eqn{uSimpleForm}{
  u^\tau = \gamma_\perp = {1 + q^2 \tau^2 + q^2 x_\perp^2 \over 2q\tau \sqrt{1+g^2}} 
    \qquad\qquad
  u^\perp = \gamma_\perp v_\perp = {q x_\perp \over \sqrt{1+g^2}} \,,
 }
where I have written out the non-zero components of $u^\mu$ in the $(\tau,\eta,x_\perp,\phi)$ coordinate system and used
 \eqn{gAgain}{
  g = {1 + q^2 x_\perp^2 - q^2 \tau^2 \over 2 q \tau} \,,
 }
which is essentially the only combination of coordinates on ${\bf R}^{3,1}$ invariant under $SO(3)_q \times SO(1,1)$.  The four-velocity \eno{uSimpleForm} might be a useful starting point for hydrodynamic simulations.  In fact, a solution to the Navier-Stokes equations can be found in closed form based on the four-velocity \eno{uSimpleForm}, provided the hydrodynamic stress tensor satisfies the constraints of conformal invariance.  Those constraints are $p = \epsilon/3$, vanishing bulk viscosity, and shear viscosity given by $\eta = {\rm H}_0 \epsilon^{3/4}$ for some dimensionless constant ${\rm H}_0$.  In this solution, the temperature in the local rest frame of the fluid is
 \eqn{HypergeometricTemperature}{
  T = {1 \over \tau f_*^{1/4}} \left( {\hat{T}_0 \over (1+g^2)^{1/3}} + 
    {{\rm H}_0 g \over \sqrt{1+g^2}} \left[ 
      1 - (1+g^2)^{1/6} {}_2F_1\left( {1 \over 2}, {1 \over 6}; {3 \over 2}; -g^2
        \right) \right] \right) \,,
 }
where $\hat{T}_0$ is a dimensionless integration constant.  Semi-realistic numbers for a central gold-gold collision at $\sqrt{s_{\rm NN}} = 200\,{\rm GeV}$ are $\hat{T}_0 = 5.55$ and ${\rm H}_0 = 0.33$, if we choose $1/q = 4.3\,{\rm fm}$.  A visual presentation of the flow can be found in figure~\ref{FlowShape}; the behavior of the energy density as a function of $\tau$ for several fixed values of $x_\perp$ can be found in figure~\ref{epsPlots}; and plots of the transverse velocity $v_\perp$ for several different values of $q$ can be found in figure~\ref{SeveralQ}.

Although my use of the $SO(3)_q$ symmetry was inspired by head-on collisions of point-sourced shocks in $AdS_5$ \cite{Gubser:2008pc}, I do not rely on strong coupling dynamics, which has been argued to lead to rapidity-dependent final states \cite{Albacete:2008vs,Albacete:2009ji}.  Instead, outside the context of the gauge-string duality, I am studying what might be termed ``conformal collisions.''  Suppose we have two shock waves of the form \eno{TuuForm} colliding head-on in flat four-dimensional Minkowski space, and suppose the underlying dynamics of the collision is a conformal field theory---but let's {\it not} make any supposition about the strength of the interactions other than to assume that at some point after the collision, hydrodynamics applies.  Finally, let's assume, as Bjorken suggested, that an approximate boost symmetry along the beam axis arises near mid-rapidity.  Then, {\it without any further knowledge of the dynamics prior to local equilibration,} we can confidently assert that the local four-velocity in the hydrodynamical phase is \eno{uSimpleForm} (near mid-rapidity, of course), and that the local temperature is \eno{HypergeometricTemperature}.  Naturally, my interest in these conformal collisions stems from the hope that central heavy-ion collisions might in some approximate sense respect $SO(3)_q$ symmetry, at least in early stages where the energy density is high enough for conformal invariance to be a good symmetry of the underlying QCD dynamics.

The $SO(3)_q$ symmetry becomes more transparent when one maps ${\bf R}^{3,1}$ conformally to its covering space $S^3 \times {\bf R}$: it is just the rotational symmetry around one particular axis through the $S^3$.  But the $SO(1,1)$ symmetry is more obscure in the $S^3 \times {\bf R}$ description.  All the symmetries are manifest in an $AdS_5$ description through the gauge-string duality.  As remarked in \cite{Gubser:2008pc,Gubser:2009sx}, off-center collisions of the light-like shocks described in section~\ref{EARLY} preserves a $U(1)$ subgroup out of the $SO(3)_q$ symmetry group of central collisions.  (In addition, some discrete symmetries are preserved.)  If the impact parameter is in the $x^2$ direction, then the $U(1)$ symmetry is the one described in \eno{ZetaOnceMore}, generalizing translations in the $x^1$ direction.  This $U(1)$ symmetry, combined with the hypothesis of boost symmetry along the beam axis, imposes significant constraints on the flow, but it seems in the non-central case that one can no longer extract the local four-velocity from symmetry considerations alone.

\section*{Acknowledgments}

I thank U.~Heinz for encouraging me to pursue the problem treated in this paper and for suggesting some improvements to an early draft.  I also thank W.~Horowitz, Y.~Kovchegov, and A.~Taliotis for a discussion, and Y.~Kovchegov for comments on the relation to collisions in $AdS_5$.  Finally, I thank A.~Yarom for discussions and for bringing reference \cite{Loganayagam:2008is} to my attention.  This work was supported in part by the Department of Energy under Grant No. DE-FG02-91ER40671 and by the NSF under award number PHY-0652782.

\appendix
\section{Another entropy estimate}

It is reasonable to assume that \eno{ChargedTracks} provides an upper bound on the entropy per unit rapidity of the fluid.  Let's consider another estimate of entropy which is more specific to the fluid.  In \cite{Adcox:2004mh} (and earlier, see e.g.~\cite{Heinz:2002un}) one can find the estimate 
 \eqn{EpsilonEstimate}{
  \epsilon = 5.4\,{{\rm GeV} \over {\rm fm}^3} \qquad\hbox{at}\qquad
   \tau=1\,{\rm fm}/c \,.
 }
A conventional assumption is that at roughly this time, the four-velocity profile of the fluid is $u = \partial_\tau$ (i.e.~boost invariant with no transverse flow), and $\epsilon(\tau,x_\perp)$ is proportional to a transverse distribution of nucleons determined by the Woods-Saxon profile:
 \eqn{TransverseWS}{
  {\epsilon(\tau,x_\perp) \over \epsilon(\tau,0)} = 
    {\displaystyle{\int_{-\infty}^\infty} {dx_3 \Big/ 
      \left[ 1 + e^{\left(\sqrt{x_3 + x_\perp^2} - R \right)/a} \right]} \over
    {\displaystyle{\int_{-\infty}^\infty} {dx_3 \Big/ 
      \left[ 1 + e^{\left(|x_3| - R \right)/a} \right]}}} \,,
 }
where $R = 6.38\,{\rm fm}$ and $a = 0.535\,{\rm fm}$.  The entropy per unit rapidity (near mid-rapidity) is
 \eqn{dSdetaWS}{
  {dS \over d\eta} = 2\pi\tau \int_0^\infty x_\perp dx_\perp \, s(\tau,x_\perp)
    = 2\pi\tau \Sigma_0 \epsilon(\tau,0)^{3/4} 
      \int_0^\infty x_\perp dx_\perp \, \left( 
        {\epsilon(\tau,x_\perp) \over \epsilon(\tau,0)} \right)^{3/4}
    \approx 3000 \,,
 }
where the final number came from plugging \eno{EpsilonEstimate}, \eno{TransverseWS}, and \eno{EntropyDensity} into the explicit integral in \eno{dSdetaWS}.  It is reassuring that the result \eno{dSdetaWS} is smaller than \eno{ChargedTracks}.

\bibliographystyle{JHEP}
\bibliography{finite}

\end{document}